\documentclass[10pt]{article}
\usepackage[textwidth=7in,textheight=8.5in]{geometry}
\usepackage{graphicx}
\usepackage{epstopdf}
\usepackage{amsmath} 
\usepackage{amssymb} 
\usepackage{fancyvrb} 
\usepackage{soul}
\usepackage{fancyhdr}
\makeatletter
\renewcommand{\maketitle}{\bgroup
\begin{flushleft}
  \begin{Huge}
  \textbf{\@title}\\
  \end{Huge}
  \vspace{1cm}
  \@author
\end{flushleft}\egroup
}
\makeatother
\title{Pion Fluctuation in High Energy Collisions - A Chaos-based Quantitative Estimation with Visibility Graph Technique}
\author{%
    \textbf{{\large Susmita Bhaduri}}$^{1}$, \textbf{{\Large Dipak Ghosh}}$^{2}$\\
    $^{1}$Deepa Ghosh Research Foundation, Kolkata-700031,India \\
    $^{2}$Deepa Ghosh Research Foundation, Kolkata-700031,India \\
    \underline{$^{1}$susmita.sbhaduri@dgfoundation.in}\\
    \underline{$^{2}$deegee111@gmail.com}
}

\date{\today}
\begin{document}

\maketitle

\begin{abstract}

We propose a new approach for studying pion fluctuation for deeper understanding of the dynamical process involved, from a perspective of fBm-based complex network analysis method called Visibility graph Analysis. This chaos-based, rigorous, non-linear technique is applied to study the erratic behavior of multipion production in \textbf{$\pi^{-}$-Ag/Br} interactions at $350$ GeV. This method can offer reliable results with finite data points. The \textbf{Power of Scale-freeness of Visibility Graph} denoted by-\textit{PSVG} is a measure of fractality, which can be used as a quantitative parameter for the assessment of the state of a chaotic system. The event-wise fluctuation of the multipion production process can be described by this parameter-\textit{PSVG}. From the analysis of the \textit{PSVG} parameter, we can quantitatively confirm that fractal behavior of the particle production process depends on the target excitation and also the fractality decreases with the increase of target excitation.
\end{abstract}


\section{Introduction}
\label{intro}
Throughout the last decade, the analysis of large density fluctuations in high energy interactions has received much attention due to its capability to provide information on the dynamics of the process of multiparticle production. A new method named intermittency was first introduced by Bialas and Peschanski~\cite{bialash986}, for the the analysis of large fluctuations. The power-law behavior of the factorial moments with respect to the size of phase-space intervals in decreasing mode, has been observed in multipion production in heavy-ion interaction. This has become indicative of self-similar fluctuation in this process. 

In the recent past, several techniques based on the fractal theory have been used to analyze the multipion emission data~\cite{hwa90,paladin1987,Grass1984,hal1986,taka1994}. Hwa(Gq moment)\cite{hwa90} and Takagi(Tq moment)\cite{taka1994} have developed the most popular of them. Considering their merits and demerits, both these methods have been extensively applied to analyze the multipion emission process~\cite{dghosh1995a,dghosh2012}. Then techniques like the Detrended Fluctuation Fnalysis(DFA) method~\cite{cpeng1994} have been used for determining monofractal scaling parameters and for detecting long-range correlations in noisy and non-stationary time-series data~\cite{MSTaqqu1995,zchen2002}. DFA method has been extended by Kantelhardt et al.~\cite{kantel2002}, for analyzing non-stationary and multifractal time-series. This generalized DFA is known as the multifractal-DFA(MF-DFA) method. 
 
If the time-series, say denoted by $x(i)$ for $i = 1,2,\ldots,N$, shows long-range(power-law) correlation, the the DFA function, say denoted by $F(s)$ follows a power-law relationship with the scale parameter, say denoted by $s$, as per the equation $F(s)\propto s^H$. This exponent $H$ is popularly known as Hurst exponent which is again related with the fractal dimension, denoted by $D_F$, as per the relationship $D_F=2-H$~\cite{kantel2001}. The extended version of DFA method called MF-DFA~\cite{kantel2002} technique has been introduced for this kind of analysis, for its advantage of having highest precision. Hurst exponent and MF-DFA parameter are used extensively in nonlinear, non-stationary analysis and they have been successful in identifying long-range correlations for different time series.
Zhang et al.\cite{YXZhang2007} applied MF-DFA method to analyze the multifractal structure of the distribution of shower particles around central rapidity region of Au-Au collisions at $\sqrt{s}_{NN}=200$A GeV. 
Multifractal analysis in particle production processes has been done in various works of recent times~\cite{Ferreiro2012,dutta2014,pmali2015}.
We have done fractal and multifractal analysis of the fluctuations of multipion production process in various types of high energy collisions\cite{dghosh1995a,dghosh2002,dghosh2005,dutta2014}, various other kinds of time series formed from natural signals like audio signals\cite{Bhaduri20162,Bhaduri20161} and biological signals like EEG and ECG signals\cite{Dutta2010,dutta2013,Ghosh2014}.
However both these methods give most accurate results for random processes like Brownian motion where the time-series has an \textit{infinite} number of data points. But in real-life situations we hardly get infinite number of data points and end up using \textit{finite} number of data points for calculation of the Hurst exponent, the \textit{MF-DFA} parameter. 
In this process the long-range correlations in the time series are fractionally broken into finite number of data points and the local dynamics relating to a particular temporal window are obviously overestimated.

In the recent past Albert and Barab{\'{a}} have reviewed the latest advances in the field of complex network and discussed the analytical tools and models for random graphs, small-world and scale-free networks~\cite{Albert2002,Barabasi2011}. Havlin et al. have discussed the application of network sciences to the description, analysis, understanding, design and repair of multi-level complex systems which occur in man-made and human social systems, in organic and inorganic matter, from nano to macro scales, and in natural and anthropogenic structures~\cite{Havlin2012}. Zhao et al. have investigated the dynamics of stock market by means of correlation-based network and identified global expansion and local clustering market behaviors during crises, using the heterogeneous time scales~\cite{Zhao20161}.
In this regard, Visibility Graph analysis\cite{laca2008,laca2009} method has gained importance due its entirely different, rigorous approach. Lacasa et al. have used fractional Brownian motion(fBm) and fractional Gaussian noises(fGn) series as a theoretical framework to analyse real-time series in different scientific fields. The Hurst parameter calculated for fBm with different methods, often yields ambiguous results, because of the presence of inherent non-stationarity and long-range dependence in fBm. Lacasa et al. applied classic method of complex network analysis to quantify long-range dependence and fractality of a time series~\cite{laca2009} and mapped fBm and fGn series into a scale-free Visibility Graph having the degree distribution as a function of the Hurst exponent~\cite{laca2009}. Zhao et al. have applied both MF-DFA and Visibility graph method to investigate
the fluctuation and geometrical structures of magnetization time series and confirmed that Hurst exponent is a good indicator of phase transition for a complex system~\cite{zhao2016}.
Visibility Graph analysis is altogether a new concept to estimate fractality from a new perspective without estimating multifractality. 
Moreover this method has recently been applied widely over time series with \textit{finite} number of data points, even with $400$ data points\cite{jiang2013}, and has achieved reliable result in various fields of science. 
This method has been used productively for analyzing various biological signals in recent works \cite{Bhaduri2014,Bhaduri20163,nil2016,bhaduriJneuro2016}. 

Target protons, also known as grey tracks as per the terminology of nuclear emulsion, are the low-energy part of intra-cascade formed from high-energy interactions. It should be noted that the number of grey particles, normally denoted by $n_g$, provides an indirect measurement of the impact or collision centrality. This centrality increases with the count of grey particles. Generally speaking, $n_g$ along can be considered as the measure of violence of target fragmentation~\cite{ander1978}. So it would be interesting to analyze the behavior of pion with respect to $n_g$ to gather more information about the inner dynamics of the particle production process in high-energy nuclear collision.
The behavior of pion with respect to $n_g$ or the number of target fragments provides more insight about the chaotic behavior of the pions in multipion production process.
In this work we have applied Visibility graph analysis to study the fractal behavior of multipion production in \textbf{$\pi^{-}$-Ag/Br} interactions at $350$ GeV with respect to target excitation. As finite number of events are available here, use of Visibility graph technique is justified.

The rest of the paper is organized as follows. The method of Visibility graph technique is presented in Section~\ref{ana}. The details of data, our analysis and the inferences from the test results are given in Section~\ref{exp}. The paper is concluded in Section~\ref{con}.

\section{Method of analysis}

We would briefly describe the Visibility graph technique in this section. 

\subsection{Visibility Graph Algorithm}
\label{ana}

\begin{figure}
\centering
\includegraphics[width=0.5\textwidth]{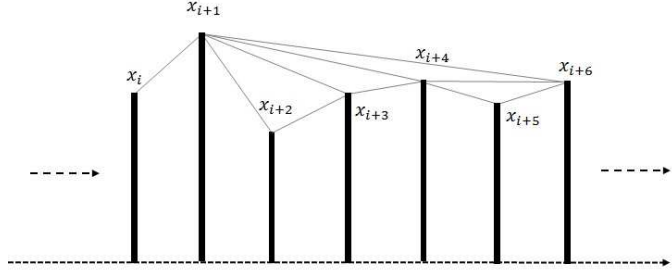}
\caption{Visibility Graph for time series X}
\label{visi}
\end{figure}
The visibility graph algorithm maps time-series $X$ to its Visibility graph. Suppose the vertex of $i^th$ point of the time series, is denoted by $X_{i}$. Two vertices (nodes) of the graph,$X_{m}$ and $X_{n}$, are said to be connected via a bidirectional edge if and only if the below equation is valid.
\setcounter{equation}{0}
\begin{equation}
X_{m+j} < X_{n} + (\frac{n - (m+j)}{n-m})\cdot(X_m - X_n) 
\label{ve}
\end{equation}
\begin{math}
\mbox{where }
\forall j \in Z^{+} \mbox{ and } j < (n-m)\\
\\
\end{math} 
In Fig.\ref{visi} it is shown that $X_{m}$ and  $X_{n}$ can see each other if the Eq. \ref{ve} is valid.
As per the Visibility graph algorithm two sequential points of the time series can see each other hence all sequential nodes are connected together.

\subsubsection{Power of Scale-freeness of VG \mbox{-} PSVG}
\label{psvg}
Lacasa et al.\cite{laca2008,laca2009} have confirmed that a fractal time series can be converted to a scale-free graph using Visibility Graph method.
The degree of a node in a graph \mbox{-} here Visibility graph, is the number of connections or edges the node has, with rest of the nodes in the graph. The degree distribution $P(k)$ of a network is then defined as the fraction of nodes with degree $k$, present in the network. Hence if there are $n$ nodes in total in a network and $n_k$ of them have degree $k$, we have $P(k) = n_k/n$ for all values of $k$.
As per Lacasa et al.\cite{laca2008,laca2009} and Ahmadlou et al.\cite{meh2012}, the degree of scale-freeness of a Visibility Graph corresponds to the amount of fractality and complexity of the time series.
According to the scale-freeness property of Visibility Graph, the degree distribution of its nodes follow power-law, i,e, $P(k) \sim k^{-\lambda_p}$, where $\lambda_p$ is a constant and it is called the \textit{Power of the Scale-freeness in Visibility Graph-PSVG}. 
Hence $\lambda_p$ or the \textit{Power of the Scale-freeness in Visibility Graph-PSVG} corresponds to the amount of self-similarity, fractality and a measure of complexity of the time series.

As the fractal dimension measures the amount of self-similarity of a time series, $\lambda_p$ is calculated from the slope of $log_2[P(k)]$ versus $log_2[1/k]$ of the time series, indicates the FD \mbox{-} Fractal Dimension of the signal\cite{laca2008,laca2009,meh2012}. 
It is observed that there is a linear relationship between $\lambda_p$ and $H$ of the associated time series\cite{laca2009}. 

Note:A power-law is a functional relationship between two quantities, where one quantity varies as a power of another.

\section{Experimental details}

\subsection{Data description}
\label{data}
Illford G5 emulsion plates were exposed to a $\pi^{-}$-beam of $350$ GeV incident energy from CERN, and the data used in this experiment was obtained from there. To scan the plates, a LeitzMetaloplan microscope with a specification of $10X$objective lens and $10X$ocular lens equipped with a semi-automatic scanning stage was used. To minimize the biases in detection, counting and measurement, each plate was scanned by two independent observers and consequently the scanning efficiency could be increased. An oil immersion-$100X$objective was used for doing the final measurement. The measuring system was integrated with both the microscopic systems having specification of $1\mu m$ resolution along $X$ and $Y$-axes, and $0.5\mu m$ resolution along $Z$-axis.

Events were selected according to the below criteria. 
\begin{itemize}
\item The incident beam-track had to lie within 3$^{\circ}$ from the axis of the main beam in the pellicle. This criteria was to ensure the selection of real projectile beam.
\item The events having the set of interactions within the range of $20\mu m$ from top and bottom surfaces of the pellicle were rejected. This helped in reducing the losses of tracks and minimizing the errors in the measurements of both emission and azimuthal angles.
\item To ensure that the events chosen shouldn't include interactions from the secondary tracks of the other interactions, all the primary beam-tracks were traced along the backward direction
\end{itemize}

The details of the events are also elaborated in our earlier works~\cite{ghosh1994,dghosh1995a,dghosh1999,ghosh2001,ghosh2002,Ghosh2011,dghosh2012}.
As per the terminology of nuclear emulsion~\cite{pow1959},the particles emitted after interactions can be classified as the shower, gray and black particles. The details of these particles are mentioned below.
%

\begin{enumerate}
\item \textit{Shower particles:} $I_{o}$ is the minimum ionization of a singly charged particle. The tracks of particles having ionization less than or equal to $1.4I_{o}$ are called shower tracks. Pions with a small admixture of K-mesons and fast protons mostly generate the shower tracks. 
The velocities of these particles are greater than $0.7c$, where $c$ is the velocity of light in free space. 

\item \textit{Grey particle:} Knocked out protons in the energy range ($30-400$ MeV), slow pions having energy of about ($30-60$ MeV) and admixture of deuterons and tritions generally produce the grey particles. They have ionization lying between $1.4I_{o}$ and $10I_{o}$. These grey particles have ranges greater than $3$mm in the emulsion medium and have velocities between $0.3c$ and $0.7c$.

\item \textit{Black particles:} They are also known as target fragments, consisting of both singly charged and multiply charged fragments. They are fragments of various elements such as carbon,lithium and beryllium, etc. with ionization greater than or equal to $10I_{o}$. 
These particles have maximum ionizing power, and are less energetic and therefore short ranged. Their ranges are less than $3$mm in the emulsion medium. Their velocities are less than $0.3c$.
\end{enumerate}

\subsection{Our method of analysis}
\label{exp}
We have divided the total number of events in three ranges of $n_g$. In doing so, although the number of events in some is comparatively low but this does not affect the result of the analysis as we have highlighted earlier in the text that this new method only can deliver reliable results with short data even with $400$ data points~\cite{jiang2013}.

To analyze the fractal behavior of pions on target excitation, we have chosen the parameter pseudorapidity-$\eta$ of produced pions in \textbf{$\pi^{-}$-Ag/Br} interactions at $350$ GeV and grouped the data as per the below three ranges of number of grey particles denoted by $n_g$. For example first dataset includes $\eta$-values of the events having number of grey particles or $n_g$ in the range $[0,2]$ or $0<=n_g<=2$. These ranges correspond to different degree of target excitation. 

\begin{enumerate}
\item $0<=n_g<=2$
\item $3<=n_g<=5$
\item $6<=n_g<=13$
\end{enumerate}

Then we have constructed Visibility graphs from the three datasets of $\eta$-values corresponding to the above three ranges of $n_g$, as per the method described in Section.~\ref{ana}. 
For each of the $3$ Visibility Graphs the following parameters are extracted.

\begin{enumerate}
\item \textbf{Heterogeneity index:}Heterogeneity indexes are extracted fore all three Visibility Graphs as per the method proposed by Estrada~\cite{Estrada2010} and listed in Table~\ref{test}. It is evident from the values that all the $3$ Visibility graphs are moderately heterogeneous. Moreover the range of values of the indexes conforms to the scale-free property of the Visibility Graphs.

\item \textbf{PSVG-$\lambda_{p}$:}The values of $k$ vs $P(k)$ are calculated for the Visibility Graphs corresponding to each of the three datasets. 
The $k$ vs $P(k)$-plot for the dataset of $\eta$-values of the events having range of $n_g$ as $0<=n_g<=2$, is shown in Fig~\ref{power}, and the power law relationship is evident here.
\begin{figure*}[t]
\centering
\includegraphics[width=0.6\textwidth]{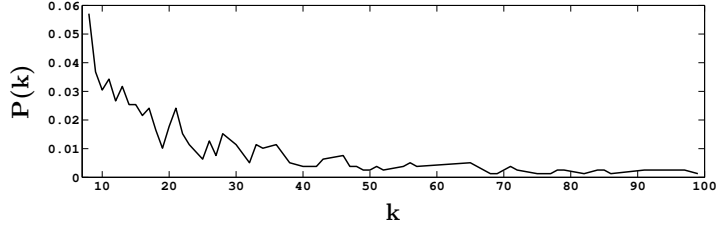}
\caption{$k$ vs $P(k)$ for the Visibility Graph created for $\eta$-values for the range $0<=n_g<=2$ for $\pi^{-}$-AgBr interaction at $350$ GeV.}
\label{power}
\end{figure*}

Power of Scale-freeness in Visibility Graph(PSVG)-$\lambda_p$, is calculated from the slope of $log_{2}[1/k]$ versus $log_{2}[P(k)]$ for each set as per the method in Section.~\ref{psvg}. Plot of $log_{2}[1/k]$ versus $log_{2}[P(k)]$ for the the range $0<=n_g<=2$ is shown in Fig~\ref{fit}. $\lambda_p$ for this range is calculated as $1.37\pm 0.09$.

\begin{figure*}[h]
\centering
\includegraphics[width=0.6\textwidth]{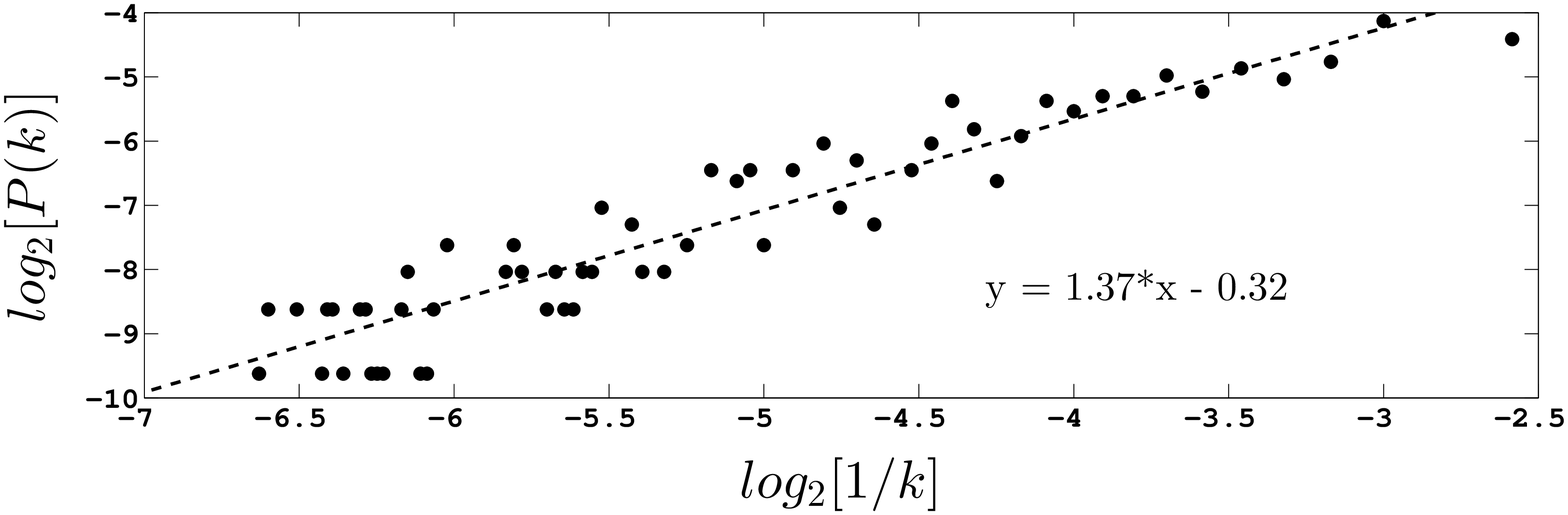}
\caption{Slope of $log_{2}[1/k]$ versus $log_{2}[P(k)]$ for the Visibility Graph created for $\eta$-values for the range $0<=n_g<=2$ for $\pi^{-}$-AgBr interaction at $350$ GeV.}
\label{fit}
\end{figure*}

For each of the Visibility graphs constructed for the datasets of $eta$-values corresponding to $3$ ranges of $n_g$, the $\lambda_{p}$-values are calculated and listed in Table~\ref{lamda_ng}.

\begin{table}[h]
\centering
\caption{$\lambda_{p}$-values calculated for $n_g$-range-wise datasets of $eta$-values.}
\label{lamda_ng}
\begin{tabular}{|c|c|c|} 
\hline
\textbf{$n_g$-range}&\textbf{$\lambda_{p}$}&\textbf{$\chi^2$/DOF}\\
\hline
$0<=n_g<=2$&$1.37\pm 0.09$&$0.02$\\
\hline
$3<=n_g<=5$&$1.23\pm 0.08$&$0.02$\\
\hline
$6<=n_g<=13$&$1.18\pm 0.08$&$0.02$\\
\hline
\end{tabular} 
\end{table}
 
The errors shown in Table~\ref{lamda_ng} are purely statistical. In our experiment, the $\chi^2$/DOF values are obtained after considering the statistical errors.

\item \textbf{Average clustering co-efficient:} Clustering co-efficient of a graph is the measurement of degree towards which nodes of the graph tend to cluster together. Average clustering co-efficients and degree for the Visibility graphs are calculated as per the method prescribed by Watts and Strogatz~\cite{Watts1998} and listed in Table~\ref{test}. Average clustering co-efficient is monotonically decreasing and average degree in increasing with increasing target excitation.

\item \textbf{Average degree:} Comparison of the average degree of the Visibility Graphs formed from each dataset is shown in the Table.~\ref{test}. It is evident that average degree is monotonically increasing with increasing target excitation.

\item \textbf{Average shortest path:} The average shortest paths between the nodes of the Visibility graphs, are calculated as per the method proposed by Johnson~\cite{johnson1977} and listed in Table~\ref{test}. This parameter is also increasing with increasing target excitation.

\item \textbf{Assortativity co-efficient:} The assortativity coefficient of the Visibility Graph is the measure of correlation of degree between pairs of linked nodes. Assortativity co-efficients for the Visibility graphs are calculated as per the method proposed by Newman~\cite{Newman2002} and listed in Table~\ref{test}. It is evident that all the $3$ graphs are disassortative. Also Assortativity co-efficients remains same for the first two ranges and decreases in the third range. 
\end{enumerate}

\begin{table*}
\centering
\caption{Trend of Heterogeneity Index, Average clustering co-efficient, Average degree, Average degree, Average shortest path and Assortativity co-eff.  for the Visibility Graphs created from the three datasets of $\eta$-values corresponding to the three ranges of $n_g$}
\label{test}
\begin{tabular}{|c|c|c|c|c|c|}
\hline
\textbf{$n_g$-range}&\textbf{Het.Index}&\textbf{Avg.Clust.co-eff.}&\textbf{Avg.Deg.}&\textbf{Avg.Shrt.Path}&\textbf{Ass.co-eff.}\\
\hline
$0<=n_g<=2$&$0.23$&$0.64$&$50.16$&$3.03$&$-0.21$\\\hline
$3<=n_g<=5$&$0.26$&$0.63$&$51.97$&$2.73$&$-0.21$\\\hline
$6<=n_g<=13$&$0.24$&$0.61$&$58.80$&$2.74$&$-0.27$\\\hline
\end{tabular}
\end{table*}

\section{Conclusion} 
\label{con}
\begin{itemize}

\item We have presented a chaos-based rigorous non-linear technique called Visibility Graph Analysis(utilizing the parameters - Heterogeneity index, Average clustering co-efficient, degree and shortest path, Assortativity co-efficient and PSVG-$\lambda_{p}$), to study the fluctuation of pions in high energy collision. In this analysis we have studied \textbf{$\pi^{-}$-Ag/Br} interactions at $350$ GeV using Visibility Graph analysis.


\item This work presents new data on scaling behavior of multiplicity fluctuations from a new perspective. As evident from Figures and Tables, the study clearly indicates that the pion multiplicity fluctuations obey a scaling law. Further we have found a decreasing trend of PSVG values($\lambda_{p}$) for three ranges of grey particles($n_g$) confirming that the fractal behavior of pion production decreases with increase of target excitation. This is an interesting and useful finding in case of hadron-nucleus interaction at high energy.

\item This study of scaling behavior from a entirely new perspective is a first of the kind analysis in the domain of high energy pionisation process yielding interesting, reliable results useful for understanding the dynamics of pion production in high energy hadronic and nuclear interaction.

\item Similar analysis can be done with the hadronic data of high energy and result would be of immense importance for modeling pionisation process in high energy nuclear collision. 
Eventually we emphasize the assessment of phase transition in high energy collision, with the help of network analysis, exploiting the PSVG(Power of the Scale-freeness in Visibility Graph, which is implicitly connected with Hurst exponent. Hurst exponent has recently been confirmed as a good indicator of phase transition for magnetization time series\cite{zhao2016}). This analysis can be extended with ALICE data of Pb-Pb collision and in future experiments with high energies to capture the onset of QGP. 
\end{itemize}


\section{Acknowledgement} 
\label{ack}
We thank the \textbf{Department of Higher Education, Govt. of West Bengal, India} for logistics support of computational analysis.

\section{References}
\providecommand{\href}[2]{#2}\begingroup\raggedright\endgroup

\end{document}